\documentclass[cits]{PoS}
\usepackage{graphicx}
 
\def\BB{{\cal B}}

\def\FF{{\cal F}}

\def\JJ{{\cal J}}
\def\KK{{\cal K}}

\def\MM{{\cal M}}
\def\NN{{\cal N}}
\def\OO{{\cal O}}

\def\QQ{{\mathbb Q}}

\def\TT{{\cal T}}

\def\WW{{\cal W}}

\def\d{{\partial}}

\newcommand{\R}{\mathbb{R}}

\newcommand{\beq}{\begin{equation}}
\newcommand{\eeq}{\end{equation}}
\newcommand{\ben}{\begin{displaymath}}
\newcommand{\een}{\end{displaymath}}
\newcommand{\beqa}{\begin{eqnarray}}
\newcommand{\eeqa}{\end{eqnarray}}
\newcommand{\bea}{\begin{eqnarray}}
\newcommand{\eea}{\end{eqnarray}}
\newcommand{\bean}{\begin{eqnarray*}}
\newcommand{\eean}{\end{eqnarray*}}
\newcommand{\ba}{\begin{array}}
\newcommand{\ea}{\end{array}}
\newcommand{\bi}{\begin{itemize}}
\newcommand{\ei}{\end{itemize}}

\title{The black M2-M5 ring intersection spins}

\ShortTitle{The black M2-M5 ring intersection spins}

\author{\speaker{Vasilis Niarchos},\thanks{CCTP-2013-02
}\\
Crete Center for Theoretical Physics, Department of Physics, University of Crete, 71303, Greece.\\
E-mail: \email{niarchos@physics.uoc.gr}}

\author{Konstadinos Siampos,\thanks{CPHT-RR027.0612
}\\
Universit\'e de Mons-Hainaut (UMH), Place du Parc, 20, B-7000 Mons, Belgium.\\
CPhT -- Ecole Polytechnique, CNRS UMR 7644, 91128 Palaiseau Cedex, France.\\
E-mail: \email{konstantinos.siampos@umons.ac.be}}

\abstract{
\vspace{0.7cm}
We review the blackfold description of the fully localized orthogonal M2-M5 intersection in 
eleven-dimensional supergravity which constitutes the direct gravitational analogue of the 
Howe-Lambert-West self-dual string soliton solution. We emphasize the preliminary implications of this 
description for the physics of the M2-M5 system and the novel research directions that it opens. 
This report contains several new results. As an extension of previous work we report new results on 
deformations of the M2-M5-KKW intersection that describe self-dual ring intersections. 
Extremal non-supersymmetric configurations of this type probe novel aspects of the M2-M5 system 
and allow us to explore further the merits and limitations of the blackfold description of black brane 
intersections. In particular, we provide an example of a situation where an effective description of 
extremal {\it non-supersymmetric} spike configurations is shown to capture correctly features of the exact 
solutions beyond its strict regime of validity. This observation hints at the existence of improved 
convergence properties in the underlying derivative expansion scheme without the necessary presence of 
supersymmetry.
}

\FullConference{Proceedings of the Corfu Summer Institute 2012 "School and Workshops on Elementary Particle Physics and Gravity"\\
		September 8-27, 2012\\
		Corfu, Greece}
		
\begin{document}

\tableofcontents

\section{Effective descriptions of the orthogonal M2-M5 intersection}

It is widely believed that much can be learned about the structure of M-theory and the non-perturbative
structure of string theory by studying the properties of the theories that reside on M2 and M5 branes.
In recent years considerable progress has been achieved by identifying the low-energy theory on 
M2 branes as an appropriately supersymmetric Chern-Simons-Matter theory 
\cite{Bagger:2007jr,Gustavsson:2007vu,Aharony:2008ug}. Our understanding of the corresponding 
question for M5 branes is more rudimentary and currently the subject of active research.

Since M2 branes can end on M5 branes it has long been suspected that the theory that resides on 
M5 branes is a mysterious novel theory of non-critical strings. These strings are charged under the 
self-dual three-form field on the M5 brane ---hence the usual reference to them as self-dual strings.
Aspects of this theory are implicit in the orthogonal M2-M5 intersection
\vspace{-0.3cm}
\beq\label{introaa}
\begin{tabular}{c c c c c c c c c c c c}
 &   &   &    &   &   &    &    &   &  &  &    \\
 \vspace{0.2cm}
 &  0 & 1  & 2   & 3  & 4  & 5   & 6   & 7  & 8 & 9 & 10 \\ 
\vspace{0.2cm}
M2~:~  & $\bullet$ & $\bullet$ & & & &  & $\bullet$ &&&&    \\
M5~:~  & $\bullet$  & $\bullet$ & $\bullet$ & $\bullet$ & $\bullet$ & $\bullet$ & & & & &   \\
\end{tabular}
\eeq
The low-energy theory at the intersection is a two-dimensional field theory with large $\NN=(4,4)$ 
superconformal symmetry. It is desirable to identify the precise properties of this theory,
for example, its central charge $c$ as a function of the number $N_2$ of M2 branes and $N_5$ of
M5 branes. Assuming unitarity the restrictive nature of the large $\NN=(4,4)$ superconformal algebra
\cite{Sevrin:1988ew} dictates that there are two integers $k_+$ and $k_-$ in terms of which
\beq\label{introab}
c=\frac{6 k_+ k_-}{k_++k_-}
~.
\eeq
It is currently unknown how $k_+$ and $k_-$ relate to $N_2$ and $N_5$.

The existence of a gravitational solution in eleven-dimensional supergravity that describes the above 
$1/4$-BPS intersection implies that there is a dual holographic description of the above
superconformal field theory (SCFT) in terms of gravity on an appropriately warped 
$AdS_3 \times S^3 \times S^3 \times \MM_2$ background with fluxes whose strength is related to 
the integers $k_+$, $k_-$. The precise form of this solution in supergravity is also unknown.
More generally, the holographic correspondence for two-dimensional SCFTs with large $\NN=(4,4)$ 
superconformal symmetry is an interesting but largely open subject (see 
\cite{de Boer:1999rh,Gukov:2004ym} and references therein for related previous work).

\subsection{Howe-Lambert-West: an effective non-gravitational description}

One of the first successful descriptions of the intersection (\ref{introaa}) 
was given by Howe, Lambert and West in \cite{Howe:1997ue} 
using the effective fivebrane worldvolume theory of a single M5 brane 
\cite{Howe:1997fb,Bandos:1997ui,Aganagic:1997zq}. In this language, a supersymmetric 
soliton solution describes how the orthogonal stack of M2 branes deforms the fivebrane worldvolume. 
The solution, which preserves the requisite $SO(1,1)\times SO(4)\times SO(4)$ symmetry,
has a non-trivial worldvolume self-dual three-form flux and a non-trivial transverse scalar field
$z=x^6$ with the profile
\beq\label{introac}
z(\sigma) = \frac{2Q_{sd}}{\sigma^2}
~.
\eeq
$\sigma$ denotes the radial distance in the directions $(2345)$ transverse to the self-dual string along
the fivebrane worldvolume. $Q_{sd}$, which is proportional to the number of M2 branes $N_2$, 
denotes the self-dual electric/magnetic charge of the self-dual string.

This solution is an M-theory analogue of the BIon solution in string theory \cite{Callan:1997kz} that 
describes how strings end on D-branes. It is also a nice illustration of the power of effective descriptions in 
string/M-theory. Although effective actions are only restricted to the description of slowly-varying 
configurations, they frequently allow us to go much further and capture non-trivial features of the theory 
bypassing the absence (in most cases) of a precise knowledge of the underlying microscopics. 

In the case of the Howe, Lambert and West solution, the above effective description 
is known currently only for the case of a {\it single} M5 brane and an arbitrary finite number of M2 branes.
The analogous description for a general number $N_5$ of M5 branes
requires knowledge of the non-abelian structure of the effective M5 brane worldvolume theory, which
is another interesting open problem.

\subsection{Blackfold funnels: an effective gravitational description}

Supergravity provides a complementary holographic description of the system in the deep 
non-abelian regime. The main technical issue in this approach is the complexity of the 
corresponding black brane intersection. There has been considerable work in this direction 
(see for instance \cite{Duff:1995yh,Tseytlin:1996as,Gauntlett:1997cv,Smith:2002wn}),
with the state-of-the-art reported in \cite{Lunin:2007mj}, but the exact fully localized $1/4$-BPS 
supergravity solution remains unknown. Non-extremal solutions are even harder to get and even 
further away from the reach of the currently known exact solution generating techniques. 

Once again effective descriptions can sidestep the complexity of the exact equations and provide a
useful new guide to the underlying physics. In the case at hand, we would be looking for an effective
treatment of black brane solutions in supergravity that describe appropriately small deformations of 
the {\it planar} M2-M5 bound state black brane solution. The latter is an exactly known supergravity
solution (see eqs. (\ref{planaraa})-(\ref{planarad}) below) that describes a stack of planar M5 branes
with dissolved M2 brane charge. The deformations of interest would describe how a black M2 brane 
solution emerges orthogonally from such a fivebrane to form the direct gravitational analogue of the 
self-dual string soliton of \cite{Howe:1997ue}. The blackfold formalism 
\cite{Emparan:2009cs,Emparan:2009at,Emparan:2011hg}, 
which provides a general effective worldvolume description of black brane dynamics, is perfectly suited for 
this task. In the case of the fully localized black M2-M5 intersection, blackfolds allow to replace the 
complicated system of partial differential equations of the exact supergravity equations of motion with a 
simpler effective fivebrane worldvolume description, which is part of a derivative expansion scheme. This
is both conceptually and computationally attractive and makes a direct comparison with the abelian 
non-gravitational self-dual string soliton solution of \cite{Howe:1997ue} much more transparent.

The details of this description for the fully localized black M2-M5 intersection were provided 
recently in \cite{Niarchos:2012pn,Niarchos:2012cy}. The precise technical issues involved in this exercise 
will be explained in greater detail below for an interesting new extension that involves stationary  
M2-M5-KKW intersections. In the rest of this introduction we proceed to highlight the main results and 
physical implications of the recent work \cite{Niarchos:2012pn,Niarchos:2012cy}.

\vspace{0.3cm}
\noindent
{\bf A novel expansion.}
The blackfold description of the M2-M5 intersection is based on a derivative expansion scheme. 
As a description of a supergravity solution it is applicable in a large-$N$ limit where $N_2, N_5 \gg 1$. 
The formalism picks out naturally two dimensionful parameters to characterize the system
\beq\label{introbaa}
q_2 \propto G Q_2~, ~~ q_5 \propto G Q_5
\eeq
where $G$ is the eleven-dimensional gravitational constant and $Q_2$, $Q_5$ the M2 and M5 charge
densities respectively (see eqs.\ (\ref{eqsag}) for the precise definitions).\footnote{A related fact is that
the near-horizon $AdS_7$ radius for a stack of M5 branes is $R_{AdS_7}=2 |q_5|^{\frac{1}{3}}$ and
the near-horizon $AdS_4$ radius for a stack of M2 branes is $R_{AdS_4}=|q_2|^{\frac{1}{6}}/\sqrt 2$.} 
The perturbative expansion that organizes the blackfold description of the intersection is controlled by the 
unique dimensionless ratio of the parameters (\ref{introbaa})
\beq\label{introba}
\frac{1}{\lambda}: = \left | \frac{q_2}{q_5^2} \right| =\frac{4N_2}{N_5^2}
~.
\eeq
This ratio controls how strongly the orthogonal M2s bend the fivebrane worldvolume. The derivative 
corrections are suppressed when $\lambda$ is large. Already at this level this picture suggests something
non-trivial about the $d=2$ SCFT at the intersection. It suggests that it admits a corresponding
large-$N$ 't Hooft-like expansion where the integers $N_2, N_5$ are scaled
to infinity with the ratio $\lambda$ kept fixed. How this expansion arises in 
quantum field theory terms is an interesting open question.

\vspace{0.3cm}
\noindent
{\bf Supersymmetric funnels.}
The effective fivebrane worldvolume theory of the black M2-M5 system involves (among other 
things) a set of five transverse scalars; the same set that describes the transverse space fluctuations of 
a fivebrane in the non-gravitational description. Specializing to extremal static configurations 
Ref.\ \cite{Niarchos:2012pn} showed that there is a non-trivial solution where a single scalar field
$z=x^6$ is turned on and behaves as
\beq\label{introbb}
z(\sigma)=2\pi \frac{N_2}{N_5} \frac{\ell_P^3}{\sigma^2}
\eeq
in direct analogy to the non-gravitational result (\ref{introac}), where $\ell_P$ is the eleven-dimensional 
Planck length scale. This spike solution is a three-sphere funnel electrically and magnetically charged 
under the three-form potential $C_3$ of eleven-dimensional supergravity. When extrapolated to the deep 
core at $\sigma=0$ it captures correctly the {\it exact} tension and charge of an extremal black M2 brane 
solution. 

This agreement with exact expectations here is at first sight rather impressive and miraculous.
At $\sigma=0$ the solution of the leading order equations of the effective description (that led to 
(\ref{introbb})) are developing very large gradients and lie well outside their strict regime of validity. The fact 
that the exact features of the emerging M2 are reproduced correctly hints that the convergence of the 
underlying expansion scheme of the effective description is much better than naively expected. This fact has 
also been stressed in the context of non-gravitional effective actions, most notably the case of the BIon 
solution \cite{Callan:1997kz} based on the use of the Dirac-Born-Infeld action. It is believed that the 
underlying reason for this improved behavior is supersymmetry. In the following sections, we will discover a 
new example where a similar observation extends beyond supersymmetry to extremal non-supersymmetric 
configurations. It appears that extremality (without necessarily supersymmetry) is enough to guarantee 
better convergence properties of the effective descriptions at hand. This is clearly an issue of general 
interest that deserves further study. In the ensuing, we rely on these improved convergence properties to 
motivate any further results based on leading order effective descriptions.

\vspace{0.3cm}
\noindent
{\bf Near-extremal funnels and central charge scalings.}
Further information about the microscopic properties of the M2-M5 system can be obtained by studying
the near-extremal thermodynamics of the supergravity solutions. The near-extremal behavior of the 
entropy is expected to be dominated by the infrared dynamics of the $d=2$ SCFT at the intersection,
and accordingly should behave at leading order in the small temperature limit in agreement with the 
Cardy formula as
\beq\label{introbc}
s= \frac{\pi c}{6} T
~,
\eeq
where $s$ is the entropy density and $c$ the central charge of the $d=2$ SCFT at the intersection.

In Ref.\ \cite{Niarchos:2012cy} we studied the thermal version of the spike (\ref{introbb}) and computed 
its near-extremal entropy. The leading order behavior in the small temperature limit behaves in accordance
to equation (\ref{introbc}) and when written in terms of $N_2$ and $\lambda$, or $N_5$ and $\lambda$, it 
yields
\beq\label{introbd}
c  \sim 0.04\, \frac{N_2^{\frac{3}{2}}}{\sqrt{\lambda}} + \ldots \sim 0.3\, \frac{N_5^3}{\lambda^2}+\ldots
~.
\eeq
The dots indicate subleading terms in negative powers of $N_2$ and $N_5$ respectively,
and negative powers of the dimensionless ratio $\lambda$. The approximations and assumptions
that go into the derivation of this result will be discussed in more detail within a more general system 
in the following sections. 

There are three key features of the leading order formulae (\ref{introbd}): the power of $N_2$
or $N_5$, the power of $\lambda$, and the overall numerical coefficient. The power of $N_2$ or 
$N_5$ is the most robust feature of the result because it follows simply from dimensional analysis.
The general Bekenstein-Hawking formula from gravity takes at leading order in the temperature 
expansion the form
\beq\label{introbe}
s = \frac{C(q_2, q_5)T}{G} 
\eeq
where $C$ is a constant that depends only $q_2$ and $q_5$ and has length dimension nine. Hence, when
expressed in terms of $N_2$ and $\lambda$, or in terms of $N_5$ and $\lambda$ the power of 
$N_2$ or $N_5$ is completely fixed as in (\ref{introbd}). At the same time this feature makes a direct 
connection with the $N_2^{\frac{3}{2}}$ scaling of the degrees of freedom of M2 branes and the 
corresponding $N_5^3$ scaling of the M5 brane degrees of freedom and is therefore highly suggestive of 
the `dual' M2 or M5 brane origin of the two-dimensional theory at the intersection.

The second feature, namely the precise power of the dimensionless ratio $\lambda$, cannot be fixed
by dimensional analysis and is therefore a specific result of the blackfold calculation. Assuming 
the basic premises of the computation we do not expect higher order terms to change this result.
The third feature on the other hand, $i.e.$ the overall numerical coefficient, is not robust 
and it is not unreasonable to expect that it changes in a more exact treatment of the system.

Finally, it is natural to ask how (\ref{introbd}) compares with the exact formula ({\ref{introab}).
When expressed in terms of $N_2$, $N_5$ eq.\ (\ref{introbd}) takes the form
\beq\label{introbf}
c \sim 0.6\, \frac{N_2^2}{N_5}+\ldots
~.
\eeq
Both (\ref{introab}) and (\ref{introbf}) are roughly speaking a ratio of the product of two integers over another 
integer, but without a more detailed knowledge about the precise relation between $k_+, k_-$ and 
$N_2, N_5$ it is impossible to decide conclusively if the results are indeed compatible. This and other 
features of the above results raise a number of interesting questions that require further study. We shall 
discuss them further in the concluding section.

\section{The M2-M5 blackfold toolkit}
\label{toolkit}

The blackfold formalism provides a general effective worldvolume description of the long-wavelength
dynamics of black branes. A detailed introduction to the formalism can be found in the original work
\cite{Emparan:2009cs,Emparan:2009at,Emparan:2011hg} and the reviews 
\cite{Emparan:2009zz,Emparan:2011br}.
For some of the previous applications of the formalism to black hole physics and string theory we refer 
the reader to 
\cite{Emparan:2009vd,Camps:2010br,Caldarelli:2010xz,Grignani:2010xm,Grignani:2011mr,
Grignani:2012iw} and references therein.
Static configurations of the M2-M5 intersection were treated recently in this framework in 
\cite{Niarchos:2012pn,Niarchos:2012cy}. In what follows we will extend the analysis of
\cite{Niarchos:2012pn,Niarchos:2012cy} to stationary M2-M5-KKW configurations by adding 
rotation in the direction of the intersection. Here we review the pertinent equations of the formalism and 
set up our notation.

The M2-M5 blackfold equations describe the long-wavelength dynamics of the black M2-M5 bound 
state \cite{Izquierdo:1995ms,Costa:1996re,Russo:1996if,Harmark:1999rb,Harmark:2000ff}}
\bea\label{planaraa}
ds_{11}^2=&(HD)^{-1/3}\Big[ -f dt^2+(dx^1)^2+(dx^2)^2+D\left( (dx^3)^2+(dx^4)^2+(dx^5)^2 \right)
\nonumber\\
&+H\left( f^{-1} dr^2 + r^2 d\Omega_4^2 \right) \Big]
~,
\eea
\beq\label{planarab}
C_3=-\sin\theta(H^{-1}-1) \coth\alpha \, dt\wedge dx^1 \wedge dx^2
+\tan\theta DH^{-1} dx^3\wedge dx^4\wedge dx^5
~,
\eeq
\beq\label{planarac}
C_6=\cos\theta D(H^{-1}-1)\coth\alpha \, dt\wedge dx^1\wedge \cdots \wedge dx^5
~,
\eeq
\beq\label{planaraca}
F_4=dC_3+\star_{11} dC_6
~,
\eeq
\beq\label{planarad}
H=1+\frac{r_0^3 \sinh^2\alpha}{r^3}~, ~~
f=1-\frac{r_0^3}{r^3}~, ~~ 
D^{-1}=\cos^2\theta+\sin^2\theta H^{-1}
~.
\eeq
The above exact solution of the supergravity equations describes a planar 
black fivebrane with M2 brane charge dissolved in its worldvolume along the (012) plane. 
The solution is parameterized by the constants $r_0$, $\alpha$ and $\theta$ 
which control the temperature, the M2 and the M5 brane charge. Angular momentum along
the fivebrane directions can be added with a general boost transformation. This transformation
introduces a unit-norm velocity six-vector with components $u^a$. A boost along the
intersection describes the black M2-M5-KKW solution 
\cite{Tseytlin:1996bh,Costa:1996re,Ohta:1997wp}.

The basic thermodynamic data of the solution (\ref{planaraa})-(\ref{planarad})
are captured by the following quantities \cite{Harmark:1999rb}
\bea\label{planarae}
\varepsilon=\frac{\Omega_{(4)}}{16\pi G}&r_0^3 (1+3\cosh^2 \alpha)
~,~~
\TT=\frac{3}{4\pi r_0 \cosh \alpha}~,~~
s=\frac{\Omega_{(4)}}{4 G} r_0^4 \cosh\alpha
~,
\\ \label{planaraeb}
Q_5=&\cos\theta \, Q~, ~~ \QQ_2=-\sin\theta \, Q~, ~~
Q=\frac{\Omega_{(4)}}{16\pi G} 3 r_0^3 \sinh\alpha\, \cosh\alpha
~,
\\
&\Phi_5=\cos\theta \, \Phi~, ~~ \Phi_2=-\sin\theta \, \Phi~, ~~
\Phi=\tanh\alpha
~.
\eea
$\varepsilon$ denotes the energy density, $\TT$ the temperature, $s$ the entropy density,
$Q_5$ the fivebrane charge, $\QQ_2$ the twobrane charge density, and $\Phi_5, \Phi_2$
the corresponding chemical potentials. We shall reserve the notation $Q$ for charges and
$\QQ$ for charge densities. In this notation $\QQ_5=Q_5$. The free energy of the solution is
\beq\label{planaraj}
\FF=\varepsilon-\TT s=\frac{\Omega_{(4)}}{16\pi G}r_0^3 (1+3\sinh^2 \alpha)
~.
\eeq
$\Omega_{(n)}=\frac{2\pi^{\frac{n+1}{2}}}{\Gamma(\frac{n+1}{2})}$ denotes the volume of the unit 
round $n$-sphere.

In analogy to standard treatments of D-branes in string theory we want to describe long-wavelength
deformations of this solution. For example, we want to describe how one spins and bends the planar 
bound state (\ref{planaraa})-(\ref{planarad}). This is achieved with an effective hydrodynamic worldvolume
theory that describes the dynamics of an effective fluid on a dynamical fivebrane worldvolume.
The relevant parameters of this theory are the variables $r_0,\alpha, \theta, u^a$ (which become local 
functions of the coordinates of the effective fivebrane worldvolume) with the addition of five 
transverse scalars that capture the bending of the fivebrane in its eleven-dimensional ambient space
and a unit three-form that keeps track of the local M2 brane current and its distribution inside the
fivebrane worldvolume. As in the fluid-gravity correspondence \cite{Bhattacharyya:2008jc,Hubeny:2011hd}, 
there is in principle a one-to-one map between the solutions of this effective theory and specific supergravity 
solutions. In what follows, we will describe only the solutions of the effective theory. The precise map to bulk 
supergravity solutions is an open problem and the subject of on-going work (see section \ref{conclude} for 
further comments).

The relevant equations of motion of the effective theory at leading order in the derivative expansion can be 
summarized as follows.

\subsection{Leading order equations of motion}
\label{geneqs}

The equations of motion are split naturally to {\it intrinsic equations} that refer to fluctuations parallel
to the worldvolume directions, and {\it extrinsic equations} that refer to fluctuations transverse to the 
worldvolume directions. The former constitute the hydrodynamic component of the dynamics, which
is also familiar from the fluid-gravity correspondence. The latter are the component that resembles
more closely the dynamics familiar from D-branes in string theory, which is usually captured by the
Dirac-Born-Infeld action (or its analogue for M-branes). 

\vspace{0.2cm}
\noindent
{\bf Intrinsic equations.}
The intrinsic equations are conservation equations for the stress-energy momentum tensor and 
the M2, M5 brane currents 
\beq\label{eqsaa}
D_a T^{ab}=0
~
\eeq
\beq\label{eqsab}
d\star_6J_3=0~, ~~ J_3=\QQ_2 \hat V_{(3)}
~,
\eeq
\beq\label{eqsac}
d\star_6J_6=0~, ~~ J_6=Q_5 \hat V_{(6)}
~.
\eeq
We use latin letters $a,b,\ldots=0,1,\ldots,5$ to denote the worldvolume coordinates.

The derivative $D_a$ is the covariant derivative with respect to the induced worldvolume metric
$\gamma_{ab}$. $T_{ab}$ is the stress-energy tensor
\beq\label{eqsad}
T_{ab}=\TT s \left(  u_a u_b-\frac{1}{3} \gamma_{ab} \right)-
\Phi_2 \QQ_2 \, \hat h_{ab} - \Phi_5 Q_5 \gamma_{ab}
~,
\eeq
which is derived from the thermodynamics of the planar solution (\ref{planarae}) with a general boost 
transformation. The relevant parameters, $r_0,\alpha,$ $etc.$, have been promoted to local functions of 
the worldvolume coordinates $\hat \sigma^a$ ($a=0,1,\ldots,5$). 
$\hat h_{ab}$ is a projector along the worldvolume directions of the 
dissolved M2 brane. In the special case of (\ref{planaraa})-(\ref{planarad}) $\hat h_{ab}$ 
projects along the plane $(012)$. In more general configurations, $\hat h_{ab}$ needs to be determined 
by solving the equations of motion.

In the charge conservation equations (\ref{eqsab}), (\ref{eqsac}) $\hat V_{(3)}$ denotes a 
unit volume 3-form along the directions of the M2 brane current and $\hat V_{(6)}$ the unit volume 
form of the fivebrane worldvolume. The six-current equation (\ref{eqsac}) implies trivially
\beq\label{eqsae} 
\d_a Q_5=0
\eeq 
from which we can deduce that $Q_5$ is an overall constant 
that participates passively in the dynamics. The M2 brane charge density and its distribution inside 
the fivebrane worldvolume, however, are dynamical quantities controlled by (\ref{eqsaa}), 
(\ref{eqsab}).

In terms of the number of M2 and M5 branes, $N_2, N_5$ respectively,
\beq\label{eqsaf}
Q_2=\frac{N_2}{(2\pi)^2 \ell_P^3}~, ~~ 
Q_5=\frac{N_5}{(2\pi)^5 \ell_P^6}
~.
\eeq
As already advertised it will also be useful to introduce the rescaled versions of the charges
\beq\label{eqsag}
q_2=-\frac{16\pi G}{3\Omega_{(3)}\Omega_{(4)}}Q_2=-4\pi^2 N_2 \ell_P^6~, ~~~
q_5=\frac{16\pi G}{3\Omega_{(4)}}Q_5=\pi N_5 \ell_P^3
~.
\eeq
The minus sign in the definition of $q_2$ is a convention.

\vspace{0.2cm}
\noindent
{\bf Extrinsic equations.}
The extrinsic equations can be recast into the form \cite{Carter:2000wv}
\beq\label{eqsba}
K_{ab}{}^\mu T^{ab}=0
\eeq
where $K_{ab}{}^\mu$ is the extrinsic curvature tensor \cite{Emparan:2009at}. For the system of
interest in this work explicit formulae for all the components of $K_{ab}{}^\mu$ appear in 
appendix \ref{extcurvature}.

\vspace{0.2cm}
In the following sections we solve the above system of equations to determine the complete profile 
of stationary configurations that describe rotating M2 branes ending on M5 branes.

\subsection{Thermodynamics}
\label{genthermo}

Stationary solutions entail a Killing vector $\xi$ that defines
unit-time translations at the asymptotic infinity and a spacelike Killing vector $\chi$ that generates 
rotations. We assume that the vector $\xi$ on the six-dimensional worldvolume $\WW_6$ is 
orthogonal to spacelike hypersurfaces $\BB_5$ and define the unit normal
\beq\label{eqsca}
n^a=\frac{1}{R_0} \xi^a \bigg |_{\WW_6}
~.
\eeq
The factor $R_0$ measures the local gravitational redshift between points on $\WW_6$ and 
the asymptotic infinity. In our case this factor will be trivial, namely $R_0=1$, but it can be 
non-trivial in other cases of interest, $e.g.$ for blackfolds in AdS backgrounds 
\cite{Caldarelli:2008pz,Armas:2010hz}.

Once we have a complete stationary solution of the equations described in the previous subsection
the leading order thermodynamics of the supergravity solution can be determined straightforwardly with the 
use of the following general expressions for the mass, angular momentum and entropy 
\cite{Emparan:2009at}
\beq\label{eqscb}
M=\int_{\BB_5}dV_{(5)} T_{ab} n^a \xi^b~, ~~
J=-\int_{\BB_5}dV_{(5)} T_{ab} n^a \chi^b~, ~~
S=-\int_{\BB_5}dV_{(5)} s u_a n^a
~.
\eeq
We shall compute these quantities in specific cases in the following sections.

\section{M2-M5-KKW ring intersections}
\label{ansatze}

In previous work \cite{Niarchos:2012pn,Niarchos:2012cy} we analyzed the static configuration
(\ref{introaa}) with symmetry group $SO(1,1)_{01}\times SO(4)_{2345} \times SO(4)_{789(10)}$ 
(the subscripts refer to the planes rotated by the corresponding symmetry group). In this case the 
intersection is an infinite open string. In what follows we extend the analysis
to describe a configuration that realizes a closed string intersection. A simple option, that preserves
the same symmetry group as above, is to compactify the direction $x^1$ of the ambient spacetime. 
This is a trivial choice that will not add anything particularly new to the analysis of \cite{Niarchos:2012pn}.

However, qualitatively new configurations on the flat eleven-dimensional Minkowski background arise by 
bending the extrinsic geometry of the M2 and M5 branes. We shall be looking for configurations 
that arise from the intersection of a cylindrical black M2 brane with a black M5 brane. A cylindrical black
M2 brane is not a regular stationary configuration unless it has non-zero angular momentum to balance
the gravitational attraction.\footnote{Appendix \ref{mthermo} summarizes the main features of the blackfold
description of rotating black M2 cylinders.} In this way, we are naturally led to the discussion of a 
deformation of the M2-M5-KKW intersection that describes rotating cylindrical M2s ending orthogonally on 
M5s along a rotating self-dual ring. Such configurations preserve a smaller symmetry group, at most 
$SO(1,1)\times SO(4) \times SO(3)$.

\subsection{Two candidate stationary configurations that fail}

\noindent
{\bf Cylindrical M2 on a planar M5.} 
We are asking if there is a fully localized supergravity solution that describes the orthogonal intersection of a 
rotating cylindrical black M2 brane along the directions $(0\vartheta 6)$ with a planar black M5 brane along
the plane (012345). The spatial part of the intersection is a circle along the angular direction 
$\vartheta$ inside the 2-plane $(12)$. The answer to this question is negative in the blackfold
approximation for the following reason.

Let us parametrize the ambient eleven-dimensional Minkowski metric as
\beq\label{ansaaa}
ds_{11}^2=-dt^2+d\rho^2+\rho^2 d\vartheta^2+dr^2+r^2 (d\psi^2+\sin^2\psi\, d\varphi^2)
+\sum_{i=6}^{10} (dx^i)^2
~.
\eeq
We have expressed the 2-plane $(12)$ in terms of the polar coordinates $(\rho,\vartheta)$ and the
transverse $\R^3$ on the fivebrane worldvolume in spherical coordinates $(r,\psi,\varphi)$.
Then, the embedding of interest is described by the ansatz
\bea\label{ansaab}
&t(\hat \sigma^a)=\hat \sigma^0~, ~~ \rho(\hat \sigma^a)=\hat \sigma^1~,
~~ \vartheta(\hat \sigma^a)=\hat \sigma^2~, ~~
r(\hat \sigma^a)=\hat \sigma^3 ~,~~ \psi(\hat \sigma^a)=\hat \sigma^4~, ~~ 
\varphi(\hat \sigma^a)=\hat \sigma^5~, 
\nonumber\\
&x^6(\hat \sigma^a):=z(r,\rho)~, ~~ x^i=0~~(i=7,8,9,10)
~.
\eea
We activate only one of the transverse scalars, $x^6$, which is a function of both 
$(\hat \sigma^1,\hat \sigma^3)=(\rho, r)$.

We are searching for a stationary solution of the equations (\ref{eqsaa})-(\ref{eqsac}), (\ref{eqsba}) 
with angular momentum along the direction $\vartheta$. For such solutions the worldvolume velocity field 
$u^a$ takes the form \cite{Emparan:2009at}
\beq\label{ansaac}
u=\frac{ \xi +\Omega \chi}{|\xi +\Omega \chi |} =\frac{1}{\sqrt{1-\Omega^2 \rho^2}}
\left( \frac{\d}{\d t}+\Omega \frac{\d}{\d \vartheta}\right)
\eeq
where $\xi=\frac{\d}{\d t}$, $\chi=\frac{\d}{\d \vartheta}$ are the Killing vectors of time translation 
and rotation of the previous section \ref{genthermo}. $\Omega$ is a the constant angular velocity of the 
configuration.

In order to have a solution with a well-defined real velocity field $u$ the worldvolume coordinate
$\hat \sigma^1=\rho$ must be bounded from above and be less than $1/\Omega$. This implies that the M5 
worldvolume is a disc in the (12) plane and cannot extend to $\rho=+\infty$. This configuration 
contradicts the conservation equation (\ref{eqsae}). Hence, we conclude that the ansatz (\ref{ansaab}) 
cannot satisfy the blackfold equations excluding M2-M5 black brane intersections of this form.

\vspace{0.2cm}
\noindent
{\bf Cylindrical M2 on a spherical M5.}
We can try to amend the flaws of the above configuration by compactifying the M5 brane worldvolume on 
a five-sphere. Using spherical coordinates we can parametrize the ambient eleven-dimensional
Minkowski metric as
\bea\label{ansaad}
&&ds_{11}^2=-dt^2+ dr^2 +r^2 d\Omega_5^2 + (dx^6)^2+ \sum_{j=8}^{10}(dx^j)^2~,
\nonumber\\
&&d\Omega_5^2 = d \varphi^2+\sin^2 \varphi \, d\theta^2 +\sum_{i=1}^3 \mu_i(\varphi,\theta)^2 d\phi_i^2
~.
\eea
The angles $\varphi,\theta \in \left[ 0, \frac{\pi}{2}\right)$, $\mu_i$ are three directional cosines\footnote{In
terms of $\varphi, \theta$: $\mu_1=\cos\theta\, \sin\varphi$, $\mu_2=\sin\theta\, \sin\varphi$, 
$\mu_3=\cos\varphi$.} and $\phi_i\in [0,2\pi)$ are three Cartan angles that parametrize the round
five-sphere. We are using spherical coordinates for the plane (123457) where the fivebrane has 
co-dimension one. The ansatz of interest is
\bea\label{ansaae}
&t(\hat \sigma^a)=\hat \sigma^0~, ~~ \phi_i(\hat \sigma^a)=\hat \sigma^i ~~(i=1,2,3)~, ~~
\varphi(\hat \sigma^a)=\hat \sigma^4~, ~~ \theta(\hat \sigma^a)=\hat \sigma^5~, ~~
r(\hat \sigma^a)=r(\varphi) ~,
\nonumber\\
&x^6(\hat \sigma^a):=z(\varphi)~, ~~ x^i=0~~(i=8,9,10)
\eea
and the fluid velocity can be taken in general to be proportional to the vector
\beq\label{ansaaf}
\frac{\d}{\d t} + \Omega_1 \frac{\d}{\d{\phi_1}} +\Omega_2 \frac{\d}{\d{\phi_2}}+\Omega_3 \frac{\d}{\d{\phi_3}}
~,
\eeq
where $\Omega_i$ $(i=1,2,3)$ are constant angular velocities. The intersection lies along a linear
combination of the Cartan angles.

With this ansatz one can run again the formalism to find solutions. There is no a priori sign in the formalism
that prevents the existence of such solutions, yet in all cases that we examined\footnote{The simplest
most symmetric configurations that can be explored have equal angular velocities 
$\Omega_1=\Omega_2=\Omega_3$.} the resulting solutions were found to be singular. Although we have 
not proven that all pertinent solutions of the above type are singular there are reasons to believe that this is
the case. Microscopic configurations where a stack of M2s ends on a spherical M5 are essentially 
disallowed by the Gauss law, $i.e.$ the fact that the flux of the self-dual three-form field has nowhere to go
on a five-sphere. It is natural to expect that the absence of analogous regular configurations in the 
blackfold formalism is gravity's way of disallowing such configurations.

\subsection{Rotating M2-M5 cylinders}
\label{cylindansatz}

A third option is to have a cylindrical black M2 end on a cylindrical black M5. This 
involves the following choice of coordinates and ansatz. We parametrize the 
ambient eleven-dimensional Minkowski metric as 
\beq\label{ansbaa}
ds_{11}^2=-dt^2+\rho^2 d\vartheta^2+dr^2
+r^2 (d\psi^2+\sin^2\psi ( d\varphi^2+\sin^2\varphi\, d\omega^2))+d\rho^2 + (dx^6)^2
+\sum_{i=8}^{10}(dx^i)^2
\eeq
and make the ansatz
\bea\label{ansbab}
&t(\hat \sigma^a)=\hat \sigma^0~, ~~ \vartheta(\hat \sigma^a)=\hat \sigma^1~, ~~
r(\hat \sigma^a)=\hat \sigma^2:=\sigma ~,~~ \psi(\hat \sigma^a)=\hat \sigma^3~, ~~ 
\varphi(\hat \sigma^a)=\hat \sigma^4~,  ~~ \omega(\hat \sigma^a)=\hat \sigma^5
\nonumber\\
&\rho(\hat \sigma^a)=\rho(\sigma)~, ~~ x^6(\hat \sigma^a):=z(\sigma)~, ~~ x^i=0~~(i=8,9,10)
~.
\eea
We are using polar coordinates $(\rho,\vartheta)$ for the plane $(17)$ where the fivebrane
worldvolume has co-dimension one. The angles $(\psi, \phi, \omega)$ parametrize a round
three-sphere. Compared to the ansatz (\ref{ansaab}) that involved
the activation of a single transverse scalar $z(r,\rho)$, here we are activating two transverse 
scalars $\rho(\sigma)$, $z(\sigma)$, which are functions of a single worldvolume coordinate.
The velocity field takes the same form as in eq.\ (\ref{ansaac}).

There are non-trivial solutions based on the ansatz (\ref{ansbab}). We shall discuss them 
in much detail in sections \ref{extremespikes} and \ref{thermalspikes}.

\subsection{Characteristic scales}
\label{characterscales}
 
The configurations of subsection \ref{cylindansatz} possess four characteristic scales: 
$(a)$ the transverse size of the black brane geometry (\ref{planaraa}) which is controlled by the charge 
radius \cite{Grignani:2010xm,Caldarelli:2010xz}
\beq\label{anscaa}
r_c(\sigma):=r_0(\sigma) (\sinh \alpha(\sigma)\, \cosh\alpha(\sigma))^{\frac{1}{3}}
~,
\eeq
$(b)$ the $S^3$ radius $\sigma$ of the induced geometry, and
$(c)$ two characteristic length sizes, $L_{\rm curv}^{(z)}$, $L_{\rm curv}^{(\rho)}$, of the 
extrinsic curvature defined in appendix \ref{extcurvature}. 
The perturbative blackfold expansion scheme is based on the hierarchy 
\cite{Emparan:2009at,Grignani:2010xm,Caldarelli:2010xz}
\beq\label{anscaa}
r_c(\sigma)\ll {\rm min}\left( \sigma, L_{\rm curv}^{(z)}, L_{\rm curv}^{(\rho)}\right)
~.
\eeq

Near extremality the leading behavior of these scales is set by the extremal solution that will be
analyzed in the next section. In that case the precise regime implied by the inequalities in (\ref{anscaa})
yields constraints that will be presented in subsection \ref{validity} below.

\section{Extremal rotating spikes}
\label{extremespikes}

It will be convenient to begin our survey of rotating spike solutions from the extremal, zero-temperature
case. In our setup one can show that this involves an extremal limit of the blackfold equations with a
null momentum wave along the direction $\vartheta$ of rotation. Such limits were studied in 
\cite{Emparan:2011hg} and take the form
\beq\label{extremeaa}
\sqrt{\TT s}u^a:=\sqrt \KK l^a~, ~~ \TT \to 0~, ~~ l^a l_a=0~, ~~ \KK={\rm finite}
~.
\eeq
The local temperature $\TT$ is scaled to zero while the velocity field $u^a$ becomes null. $\KK$ 
denotes the null momentum density, which will be a worldvolume-dependent quantity to be
determined by solving the equations of motion. In this limit
the stress-energy tensor $T^{ab}$ that controls the blackfold equations becomes
\beq\label{extremeab}
T^{ab}=\KK l^a l^b-\Phi_2 \QQ_2 \hat h^{ab}-\Phi_5 Q_5 \gamma^{ab}
~.
\eeq

With the ansatz of subsection \ref{cylindansatz}
\beq\label{extremeac}
l=\frac{\d}{\d t}+\Omega \frac{\d}{\d \theta}
\eeq
and the velocity field (\ref{ansaac}) becomes null when 
\beq\label{extremead}
\rho=\frac{1}{\Omega}
~.
\eeq
The induced metric $\gamma_{ab}$ and the two-brane projector $\hat h_{ab}$ take the form 
\beq\label{extremeae}
\gamma={\rm diag}\left( -1,\rho^2,1+{z'}^2,\sigma^2,\sigma^2 \sin^2\psi, \sigma^2 \sin^2\psi \, 
\sin^2\varphi \right)
~,
\eeq
\beq\label{extremeaf}
\hat h={\rm diag}\left(  -1,\rho^2,1+{z'}^2,0,0,0 \right)
~.
\eeq
We shall be using the notation $':=\frac{d}{d\sigma}$.

\subsection{The solution}

We are solving the equations 
\beq\label{extremeba}
K_{ab}{}^\mu T^{ab}=0
~,
\eeq
\beq\label{extremebb}
D_a T^{ab}=0
~,
\eeq
\beq\label{extremebc}
d\star_6(\QQ_2 \hat V_{(3)})=0
\eeq
for the unknown functions $z(\sigma)$, $\theta(\sigma)$, $\KK(\sigma)$, $Q(\sigma)$.
In the extremal limit we scale the variables $r_0$, $\alpha$ to $0$, $\infty$ respectively
keeping the quantity $Q$ in eq.\ (\ref{planaraeb}) fixed. Consequently, 
\beq\label{extremebd}
\Phi_2=-\sin\theta~,~~ \Phi_5=\cos\theta~, ~~ \Phi=1
~.
\eeq

The set of extrinsic equations (\ref{extremeba}) gives two independent equations (for explicit expressions
of the extrinsic curvature tensor see (\ref{ringav}) in appendix \ref{extcurvature}) 
\beq\label{extremebe}
\frac{z''}{z'(1+{z'}^2)}=-\frac{3\cos^2\theta}{\sigma}~, ~~~ \KK=Q ~\Rightarrow~ T^{11}=0
~.
\eeq
The second equation reproduces the condition of vanishing tension familiar from previous studies 
of black ring solutions \cite{Emparan:2007wm}.

The intrinsic equations (\ref{extremebb}), (\ref{extremebc}) lead to two additional independent relations
\beq\label{extremebf}
\sigma \KK'+3\sin^2\theta\, \KK=0
~,
\eeq
\beq\label{extremebg}
(\sigma^3 \sin \theta \, \KK)'=0
~
\eeq
where we made use of the second equation $\KK=Q$ in (\ref{extremebe}). 
We solve both of them by setting
\beq\label{extremebi}
\cos\theta=\frac{1}{\sqrt{1+\frac{\kappa^2}{\sigma^6}}}~, ~~ \KK=Q
=Q_5 \sqrt{1+\frac{\kappa^2}{\sigma^6}}
~.
\eeq
$\kappa$ is a constant which is related to the total M2, M5 charges in the following way
\beq\label{extremebj}
\kappa=\frac{1}{2\pi^2} \frac{Q_2}{Q_5}=4\pi \frac{N_2}{N_5} \ell_p^3
~.
\eeq

The three-funnel profile $z(\sigma)$ is now fixed completely by the first equation in (\ref{extremebe}),
which takes the same form as in the static case analyzed in \cite{Niarchos:2012pn}. With boundary 
conditions
\beq\label{extremebk}
\lim_{\sigma \to +\infty}z(\sigma)=0~, ~~
\lim_{\sigma \to \sigma_0^+}z'(\sigma)=-\infty
\eeq
we find
\beq\label{extremebl}
z(\sigma)=\frac{\sqrt{\sigma_0^6+\kappa^2}}{2\sigma^2} \,
_2F_1\left( \frac{1}{3},\frac{1}{2},\frac{4}{3}; \frac{\sigma_0^6}{\sigma^6}\right)
~.
\eeq
The spike solution that reproduces (\ref{introbb}) has $\sigma_0=0$.

The above solutions are zero-temperature extremal configurations. For zero angular velocity, 
$\Omega=0$, the radius $\rho$ of the intersection becomes infinite and we recover the static solutions of 
\cite{Niarchos:2012pn},\footnote{The static solutions \cite{Niarchos:2012pn} have the same 
$z$ profile as in (\ref{extremebl}).} which are expected to be 1/4-BPS for $\sigma_0=0$. As was 
mentioned previously, the addition of angular momentum breaks some of the overall symmetry.
As we discuss further in a moment we have good reasons to believe that it also breaks completely the 
supersymmetry. Hence, in the presence of non-zero angular momentum all of the above solutions are 
extremal but non-supersymmetric.

\subsection{Validity of the leading order approximation}
\label{validity}

Near extremality the leading behavior of the characteristic scales 
$r_c, L_{\rm curv}^{(z)}, L_{\rm curv}^{(\rho)}$ (see subsection \ref{characterscales})
is set by the extremal solution we have just presented. One can show, using the definitions and above 
expressions, or by taking the extremal limit of eqs.\ (\ref{thermalag})-(\ref{thermalai}) below, that 
\beq\label{anscab}
r_c(\sigma)=q_5^{\frac{1}{3}}\left( 1+\frac{\kappa^2}{\sigma^6} \right)^{\frac{1}{6}}
\eeq
with the constant $\kappa$ defined in eq.\ (\ref{extremebj}) and $q_5$ the constant in eq.\ (\ref{eqsag}).
The leading behavior of the characteristic curvature sizes $L_{\rm curv}^{(z)}$, $L_{\rm curv}^{(\rho)}$
can be determined with the use of the expressions in appendix \ref{extcurvature}. We find 
\beq\label{anscac}
L_{\rm curv}^{(z)}=\frac{\sigma(\sigma^6+\kappa^2)^{\frac{3}{2}}}{3\kappa^2\sqrt{\sigma_0^6+\kappa^2}}
~,~~ 
L_{\rm curv}^{(\rho)}=\frac{1}{\Omega}
~.
\eeq

Moreover, it was argued in Ref.\ \cite{Niarchos:2012cy} that the perturbative corrections to the leading 
order result are controlled by negative powers of the ratio $\lambda=\frac{N_5^2}{4N_2}$. It will therefore 
be natural to work in the limit $\lambda \gg 1$ where these corrections are suppressed.

Consequently, the first inequality in (\ref{anscaa}), $r_c(\sigma)\ll \sigma$, implies
\beq\label{anscad}
\sigma \gg \sigma_c=\left( \frac{\pi N_5}{\sqrt 2} \right)^{\frac{1}{3}}
\left( 1+\sqrt{1+\frac{64 N_2^2}{N_5^4}}\right)^{\frac{1}{6}} \ell_P
~\eeq
where by definition $\sigma_c$ is the point where $r_c(\sigma_c)=\sigma_c$. The second inequality,
$r_c(\sigma)\ll L_{\rm curv}^{(z)}$, is automatically satisfied given (\ref{anscad}). The third inequality,
$r_c(\sigma)\ll L_{\rm curv}^{(\rho)}$, is satisfied only when
\beq\label{anscaf}
\sigma\gg \tilde \sigma_c:=\frac{\Omega q_2^{\frac{1}{3}}}{(1-\Omega^6 q_5^2)^{\frac{1}{6}}}
~.
\eeq
We are assuming
\beq\label{anscae}
\Omega<q_5^{-\frac{1}{3}}
~,
\eeq
which places an upper bound on the angular velocity $\Omega$. Since 
\beq\label{anscag}
\tilde \sigma_c \leqslant \left( \frac{\kappa}{2} \right)^{\frac{1}{3}} <
\left(\frac{1}{\sqrt{2}\lambda}\right)^{\frac{1}{3}} \sigma_c
\eeq
we deduce that the inequality (\ref{anscaf}) is automatically satisfied in the limit $\lambda\gg 1$ given 
(\ref{anscad}).

We conclude that there is a characteristic lower bound scale $\sigma_c(N_2,N_5)$ on the radius 
$\sigma$ below which the leading order blackfold description of the extremal fully localized M2-M5 
intersection is beyond the strict regime of its validity. Any quantity computed below this scale cannot be 
trusted and is anticipated a priori to diverge considerably from the exact result. Interestingly, there are
quantities for which such deviations are not observed.

\subsection{Miraculous agreement beyond the strict regime of validity}

The mass and angular momentum densities of the solution (\ref{extremebl}) can be determined 
straightforwardly from the general expressions (\ref{eqscb}). With a little algebra one finds
\beq\label{extremeca}
\frac{dM}{dz}=2\Omega \frac{dJ}{dz}=\frac{4\pi \Omega_{(3)} \KK}{\Omega} 
\frac{\sqrt{1+{z'}^2}}{z'} \sigma^3
=-\frac{4\pi Q_2}{\kappa \Omega} \frac{\kappa^2+\sigma^6}{\sqrt{\kappa^2+\sigma_0^6}}
~.
\eeq
The entropy density of the extremal solutions is vanishing.

It is interesting to note the following property.
The spike solutions have $\sigma_0=0$ and extend all the way to $z=+\infty$. At that point, namely
the tip of the spike, the mass and angular momentum densities reproduce exactly the corresponding 
quantities of extremal cylindrical M2 black branes. Indeed, (\ref{extremeca}) gives
\beq\label{extremecb}
\frac{dM}{dz}\bigg |_{\sigma=\sigma_0=0} = -\frac{4\pi Q_2}{\Omega}=T_{M2}
~, ~~\frac{dJ}{dz} \bigg |_{\sigma=\sigma_0=0}=-\frac{2\pi Q_2}{\Omega^2}=\JJ_{M2}
\eeq
where $T_{M_2}$ and $\JJ_{M2}$ are respectively the tension and angular momentum density of an 
extremal cylindrical M2 black brane with the same charge (see eq.\ (\ref{mar}) in appendix \ref{mthermo}).

This matching is impressive because it occurs in a region $(\sigma \sim 0 \ll \sigma_c)$ 
where the leading order blackfold approximation breaks down. The same impressive matching has been 
observed also in the case of the F1-D3 BIon \cite{Grignani:2010xm}, and the case of the 
static M2-M5 intersection \cite{Niarchos:2012pn}, generalizing in gravity the corresponding observations 
in the DBI treatment of the BIon \cite{Callan:1997kz}. In both cases the extremal intersection is 
supersymmetric making supersymmetry the natural candidate for the underlying reasons behind this 
miraculous matching. This matching is an indication that under suitable conditions, $e.g.$ supersymmetry, 
effective treatments like the DBI, blackfolds $etc.$ can enjoy much better convergence properties than
naively anticipated on general grounds.

In the case at hand, however, the agreement (\ref{extremecb}) is even more impressive. Besides the 
fact that it involves the matching of two quantities (tension and angular momentum density), 
it occurs for a solution that is extremal but non-supersymmetric. Indeed, the solution describes a fivebrane
that deforms to become in a region of spacetime a rotating cylindrical M2 black brane. Since a cylindrical 
M2 brane has only a dipole M2 charge, it is not expected to be supersymmetric \cite{Emparan:2011hg}.
This implies that the whole configuration is extremal but non-supersymmetric. 
Consequently, this is an interesting new example where the cancellations (and better convergence 
properties) implied by the above matching occur at extremality without the necessary presence of 
supersymmetry. Further related unexpected agreements will be noted for the near-extremal solutions in the 
next section.

\section{Thermally excited rotating spikes}
\label{thermalspikes}

Further aspects of the system are revealed by studying it at non-zero temperature.
For instance, near-extremal black brane thermodynamics is one way to derive information about 
the microscopic structure of the system \cite{Klebanov:1996un}, and this is how we arrived at the 
central charge expression (\ref{introbd}) in Ref.\ \cite{Niarchos:2012cy} for the static $\Omega=0$
intersection. In this section we proceed to analyze the near-extremal properties of the 
rotating spike solutions generalizing the discussion of Ref.\ \cite{Niarchos:2012cy}. 
One of the new features of the finite-temperature solutions is that $\rho$ is no longer constrained by the 
extremality condition (\ref{extremead}). Hence, both transverse scalars $\rho$ and $z$ are now 
non-trivial functions of the worldvolume coordinate $\sigma$ and we have to solve a more complicated
system of differential equations.

\subsection{Equations of motion and general thermodynamic expressions}
\label{thermaleom}

A finite-temperature configuration is obtained by solving the blackfold equations 
(\ref{extremeba})-(\ref{extremebc}) for the unknown functions $z(\sigma)$, $\rho(\sigma)$, $r_0(\sigma)$,
$\alpha(\sigma)$, $\theta(\sigma)$. The induced metric and two-brane projector $\hat h_{ab}$ 
take the explicit form
\beq\label{thermalaa}
\gamma={\rm diag}\left( -1,\rho^2,1+{\rho'}^2+{z'}^2,\sigma^2,\sigma^2 \sin^2\psi, \sigma^2 \sin^2\psi \, 
\sin^2\varphi \right)
~,
\eeq
\beq\label{thermalab}
\hat h={\rm diag}\left(  -1,\rho^2,1+{z'}^2+{\rho'}^2,0,0,0 \right)
~.
\eeq
Since we are looking for stationary solutions the velocity field is oriented along a Killing vector 
\beq\label{thermalac}
u=\frac{1}{\sqrt{1-\Omega^2 \rho^2}}\left( \frac{\d}{\d t}+\Omega \frac{\d}{\d \vartheta} \right)
~\eeq
and the intrinsic equations are satisfied by requiring that the following quantities
\cite{Niarchos:2012pn,Niarchos:2012cy}
\beq\label{thermalad}
q_2=-\frac{16 \pi G}{3\Omega_{(3)}\Omega_{(4)}} Q_2
=\sigma^3 r_0^3 \sin\theta\, \sinh \alpha\, \cosh\alpha
~,
\eeq
\beq\label{thermalae}
q_5=\frac{16\pi G}{3\Omega_{(4)}}Q_5= r_0^3 \cos\theta\, \sinh \alpha\, \cosh\alpha
~,
\eeq
\beq\label{thermalaf}
\frac{r_0 \cosh\alpha}{\sqrt{1-\Omega^2 \rho^2}}=\beta:=\frac{3}{4\pi T}
~.
\eeq
are $\sigma$-independent constants. $q_2, q_5$ are the same rescaled versions of the total charges 
$Q_2, Q_5$ as in eq.\ (\ref{eqsag}), while $T$ expresses the global temperature of the solution. This 
leaves the extrinsic equations that determine the profile of the transverse scalars $z,\rho$, which will be discussed momentarily.

Solving the above equations for the unknowns $\alpha, \theta, r_0$ we find, as in \cite{Niarchos:2012pn}, 
two solutions (denoted by the subscript $\pm$)
\beq\label{thermalag}
\cosh \alpha_\pm =\frac{\beta^3}{\sqrt 2 q_5}(1-\Omega^2 \rho^2)^{\frac{3}{2}}
\frac{\sqrt{1\pm \sqrt{1-\frac{4q_5^2}{\beta^6}\left(1+\frac{\kappa^2}{\sigma^6}\right)
(1-\Omega^2\rho^2)^{-3}}}}{\sqrt{1+\frac{\kappa^2}{\sigma^6}}}
~,
\eeq
\beq\label{thermalai}
r_{0,\pm}=\frac{\sqrt 2 q_5}{\beta^2} \frac{(1-\Omega^2 \rho^2)^{-1}\sqrt{1+\frac{\kappa^2}{\sigma^6}}}
{\sqrt{1\pm \sqrt{1-\frac{4q_5^2}{\beta^6}\left(1+\frac{\kappa^2}{\sigma^6}\right)
(1-\Omega^2\rho^2)^{-3}}}}
~,
\eeq
\beq\label{thermalaj}
\cos\theta=\frac{1}{\sqrt{1+\frac{\kappa^2}{\sigma^6}}}
~.
\eeq
Notice that the above expressions break down at the critical value 
\beq\label{thermalak}
\sigma_b=\left( \frac{4q_2^2}{\beta^6(1-\Omega^2 \rho_b^2)^3-4q_5^2}\right)^{\frac{1}{6}}~, ~~
\rho_b:=\rho(\sigma_b)
~.
\eeq
The more explicit form of this value in the near-extremal regime will be discussed further in 
subsection \ref{scalehierarchies} below.

One can show \cite{Emparan:2009at,Emparan:2011hg,Niarchos:2012pn} 
that the remaining, extrinsic, equations for the transverse scalars can be obtained by extremizing the 
Dirac-like action
\beq\label{thermalal}
I=\frac{\Omega_{(3)}\Omega_{(4)}}{8G} L_t \int d\sigma\, \sqrt{1+{\rho'}^2+{z'}^2}F_\pm (\rho,\sigma)
~,
\eeq
\beq\label{thermalam}
F_\pm(\rho,\sigma)=\beta^3 \sigma^3 \rho \frac{1+3\sinh^2\alpha_\pm}{\cosh^3\alpha_\pm}
(1-\Omega^2 \rho^2)^{\frac{3}{2}}
\eeq
where the explicit form of $\alpha_\pm$ in eq.\ (\ref{thermalag}) is implied.
$L_t$ denotes the (possibly infinite) length of the time direction.
We emphasize that this action is suitable only for stationary configurations. For more generic 
deformations of the M2-M5 bound state one should consider directly the full blackfold equations
(\ref{extremeba})-(\ref{extremebc}).

Varying (\ref{thermalal}), (\ref{thermalam}) with respect to the unknown scalars $\rho,z$ we obtain the 
equations of motion
\beq\label{thermalan}
\d_\rho F_\pm \sqrt{1+{\rho'}^2+{z'}^2}=\left( \frac{\rho' F_{\pm}}{\sqrt{1+{\rho'}^2+{z'}^2}}\right)'
~,\eeq
\beq\label{thermalao}
\left( \frac{z' F_\pm}{\sqrt{1+{\rho'}^2+{z'}^2}} \right)'=0
~.
\eeq
Using the boundary conditions (\ref{extremebk}) and combining with (\ref{thermalao}) we can also recast 
(\ref{thermalan}) into the more convenient form
\beq\label{thermalap}
\left( \frac{\rho'}{z'}\right)'=z' \frac{F_\pm \d_\rho F_\pm}{F_{0,\pm}^2}~, ~~
F_{0,\pm}:=F_\pm(\rho(\sigma_0),\sigma_0)
~.
\eeq

In the next subsection we solve these complicated equations perturbatively in the small temperature
limit. Once we have a solution its thermodynamic data are immediately obtainable from the following
expressions
\beq\label{thermalaq}
\frac{dM}{dz}=\frac{\Omega_{(3)}\Omega_{(4)} \beta^3}{8G} \sigma^3 \rho 
\sqrt{1-\Omega^2\rho^2}\frac{1+3\cosh^2\alpha_\pm -\Omega^2\rho^2(1+3\sinh^2\alpha_\pm)}
{\cosh^3\alpha_\pm} \frac{F_\pm}{F_{0,\pm}}
~,
\eeq
\beq\label{thermalar}
\frac{dJ}{dz}=\frac{3\Omega_{(3)}\Omega_{(4)}}{8G} \frac{\Omega \beta^3 \sigma^3 \rho^3
\sqrt{1-\Omega^2\rho^2}}{\cosh^3\alpha_\pm} \frac{F_\pm}{F_{0,\pm}}
~,
\eeq
\beq\label{thermalas}
\frac{dS}{dz}=\frac{\pi \Omega_{(3)}\Omega_{(4)}\beta^4}{2G} \frac{\sigma^3 \rho 
(1-\Omega^2 \rho^2 )^{\frac{3}{2}}}{\cosh^3\alpha_\pm} \frac{F_\pm}{F_{0,\pm}}
~.
\eeq

\subsection{Small temperature expansion}

The equations of motion (\ref{thermalan}), (\ref{thermalao}) admit a perturbative small temperature 
($\beta \gg 1$) expansion with
\beq\label{thermalba}
z(\sigma)=\sum_{n=0}^\infty z_n(\sigma) \beta^{-\frac{3n}{2}}~, ~~
\rho(\sigma)=\sum_{n=0}^\infty \rho_n(\sigma) \beta^{-\frac{3n}{2}}
~.
\eeq
In this expansion the coefficients $\rho_n(\sigma)$ are determined order by order by algebraic 
equations, and $z_n(\sigma)$ by first order ODEs. Focusing on the $+$ branch that connects smoothly 
to the extremal solutions, we find at the first few orders 
\bea\label{thermalbb}
&\rho(\sigma)=\frac{1}{\Omega}\bigg( 1-\frac{\sqrt{q_2}}{2} \frac{(1+\frac{\sigma^6}{\kappa^2})^{\frac{1}{4}}}
{\sigma^{\frac{3}{2}}} \beta^{-\frac{3}{2}}
\nonumber\\
&-\frac{q_2}{192\kappa \Omega^2}
\frac{9\kappa^2(-8\sigma^{12}+14 \sigma^6 \sigma_0^6+\kappa^2(\sigma^6+5\sigma_0^6))
+40\sigma^2 (\kappa^2+\sigma_0^6)^3 \Omega^2}
{\sigma^5(\kappa^2+\sigma^6)^{\frac{5}{2}}} \beta^{-3}+\OO\left(\beta^{-\frac{9}{2}}\right)\bigg)
,
\eea
\beq\label{thermalbc}
z'(\sigma)=-\sqrt{\frac{\sigma_0^6+\kappa^2}{\sigma^6-\sigma_0^6}}
+\frac{2\sqrt{q_2}}{3\sqrt \kappa} \frac{(\kappa^2+\sigma^6)\sqrt{\kappa^2+\sigma_0^6}}
{(\sigma^6-\sigma_0^6)^{\frac{3}{2}}}\left[ \left(1+\frac{\kappa^2}{\sigma_0^6}\right)^{\frac{1}{4}}
-\left(1+\frac{\kappa^2}{\sigma^6} \right)^{\frac{1}{4}}\right]\beta^{-\frac{3}{2}}+\OO(\beta^{-3})
.\eeq
Integrating the last equation over $\sigma$ with the boundary condition 
$\displaystyle{\lim_{\sigma\to +\infty}z(\sigma)=0}$ we determine the precise form of the function 
$z(\sigma)$. Notice in (\ref{thermalbb}) that the behavior of the functions $(z_0',z_1')$ around 
the points $\sigma_0$, $+\infty$ is 
\beq\label{thermalbd}
(z_0',z_1') \sim \frac{1}{\sqrt{\sigma-\sigma_0}}~, ~~ {\rm as} ~~\sigma\to \sigma_0^+~~~{\rm and}~~~
(z_0',z_1') \sim \frac{1}{\sigma^3}~, ~~{\rm as}~~\sigma\to +\infty
~.
\eeq
Hence, the functions $(z_0,z_1)$ converge in $[\sigma_0,+\infty)$. The analytic expression of 
$z_0(\sigma)$ was given in formula (\ref{extremebl}). For $z_1(\sigma)$ we have not been able to find
a solution in closed analytic form, but one can be found numerically.

\subsection{Note on characteristic scale hierarchies}
\label{scalehierarchies}

So far the perturbative solutions (\ref{thermalbb}), (\ref{thermalbc}) have an undetermined parameter, $\sigma_0$. 
In order to determine the precise form of a finite-temperature rotating spike solution
$\sigma_0$ should be fixed in terms of the constant parameters $T$, $\Omega$. In the next 
subsection we describe how this can be achieved with a prescription proposed in \cite{Grignani:2011mr}. 
Before that, however, it will be useful to make a small parenthesis to note a related hierarchy of scales that 
characterize our setup.

In subsections \ref{characterscales} and \ref{validity} we discussed a characteristic radial scale 
$\sigma_c(N_2,N_5)$ (\ref{anscad}) below which the perturbative
blackfold expansion scheme breaks down. In the near-extremal regime this scale is controlled 
solely by the number of M2 and M5 branes, and is independent of the temperature and angular 
velocity $T,\Omega$.

In subsection \ref{thermaleom} we discussed another characteristic scale $\sigma_b$, (\ref{thermalak}), 
where the leading order blackfold solution breaks down. With the use of the explicit perturbative form
of the function $\rho(\sigma)$ (\ref{thermalbb}), we find that the leading order temperature dependence
of $\sigma_b$ is
\beq\label{thermalca}
\sigma_b \sim \frac{q_2^{\frac{1}{3}}}{\beta}
~.
\eeq

\begin{figure}[!t]
\begin{center}
\vskip -0.5cm
\begin{tabular}{cc}
\includegraphics[height=8.2cm]{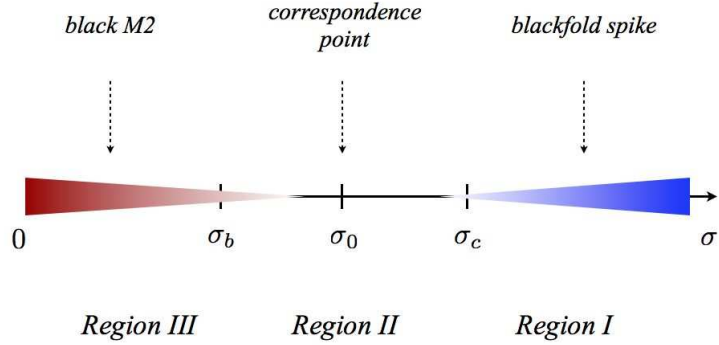}
\end{tabular}
\end{center}
\vskip -1.5 cm \caption{Three different regions (I, II, III) on the semi-infinite $\sigma$ line. 
$\sigma_c$ is the characteristic scale
where the regime of validity of the leading order blackfold approximation ends. $\sigma_b$ is the
point where a thermal spike solution breaks down. $\sigma_0$ denotes the correspondence point where
the thermodynamic data of a black M2 brane are matched to the thermodynamic data of
the blackfold spike. Figure reprinted from \cite{Niarchos:2012cy}.}
\label{sigma.values}
\end{figure}

In the next subsection we shall use the prescription of Ref.\ \cite{Grignani:2011mr} to determine $\sigma_0$. 
Anticipating the result, which exhibits the scaling behavior
\beq\label{thermalcb}
\sigma_0\sim \left( \frac{q_2}{\beta} \right)^{\frac{1}{5}}~,
\eeq
we deduce the important near-extremal hierarchy of scales
\beq\label{thermalcc}
\sigma_b \ll \sigma_0 \ll \sigma_c
\eeq
that was noted also in \cite{Niarchos:2012cy} for the static M2-M5 intersection. This behavior is depicted 
in figure \ref{sigma.values}.

\subsection{Boundary conditions}

The extremal spike that describes the rotating M2-M5 intersection is a solution 
with $\sigma_0=0$. Although the tip, at $\sigma_0=0 \ll \sigma_c$, lies well beyond the regime of 
validity of the blackfold expansion scheme, we have seen that the leading order computation 
(\ref{extremecb}) recovers correctly the tension and angular momentum density of the intersecting M2 
branes. 

It has been proposed \cite{Grignani:2011mr} (in the context of the F1-D3 system) that the near-extremal 
spike solutions can be determined by requiring that they exhibit an analogous matching of thermodynamic
data at a non-vanishing $\sigma_0$ controlled by the finite temperature. Again, this entails an 
extrapolation away from the regime of validity of the descriptions we are using whose success relies 
on a miraculous agreement with exact data that is not immediately obvious.

In Ref.\ \cite{Niarchos:2012cy} we presented favorable evidence for the applicability of this prescription
to the near-extremal static M2-M5 intersection. By matching the tension and entropy density of the
blackfold spike to the corresponding quantities of a static black M2 at $\sigma_0$ one obtains,
as in the case of the F1-D3 system \cite{Grignani:2011mr}, two independent conditions for a single 
unknown. The two resulting values of $\sigma_0$ were found to agree within 4\% at the leading order 
temperature dependence providing evidence for the consistency of the approach. 
Since there is no a priori justification for this rather useful prescription it is interesting to know if it is
equally successful in other more generic setups. The rotating M2-M5 intersection provides a new 
more involved example where instead of two independent matching conditions we have four. 
The miracles of the extremal description imply that any deviations will be suppressed by the temperature.
Hence, it is sensible to focus on the leading order temperature behavior of the solutions in 
the near-extremal limit.

The precise form of the matching conditions is the following
\beq\label{thermalda}
\left( \frac{dM}{dz}\bigg |_{\sigma=\sigma_0^+}\right)_{\rm M2-M5}=T_{M2}
~,~~
\left( \frac{dJ}{dz}\bigg |_{\sigma=\sigma_0^+}\right)_{\rm M2-M5}=\JJ_{M2}
~,
\eeq
\beq\label{thermaldc}
\left( \frac{dS}{dz}\bigg |_{\sigma=\sigma_0^+}\right)_{\rm M2-M5}=s_{M2}
~,~~
\rho(\sigma_0)_{\rm M2-M5}=\rho_{\rm M2}
\eeq
where $T_{M_2}$, $\JJ_{M_2}$, $s_{M_2}$ and $\rho_{M_2}$ denote respectively the 
tension, angular momentum density, entropy density and cylinder radius of the black M2
evaluated at the same temperature, angular velocity and number of M2 branes as in the M2-M5
system. The second condition in (\ref{thermaldc}) requires the matching of a geometric datum.
Each of these conditions determines independently a specific value for $\sigma_0$ and there is no apparent
reason why these values should be the same. The observed agreement or disagreement
is a measure of how well the matching scheme (\ref{thermalda}), (\ref{thermaldc}) works.

The matching of the two descriptions (the M2 and the deformed M5) at $\sigma_0$ is performed in an 
intermediate region where neither is strictly valid \cite{Grignani:2011mr,Niarchos:2012cy}.
From the extrapolated leading order M2-M5 blackfold description this implies that the matching is performed 
in a regime where
\beq\label{thermalde}
\sigma_0\ll \sigma_c~, ~~ \sigma_0^6 \ll \kappa^2
~.
\eeq
The first inequality declares that we are working outside the strict regime of validity of the blackfold
description, and the second inequality that we are describing a region where the contribution of the
M2 branes dominates. With these assumptions the small temperature expansion of the quantities 
appearing in (\ref{thermalda}), (\ref{thermaldc}) yields the following results.

From the expansion of the tensions we find
\bea\label{thermaldf}
\left( \frac{dM}{dz}\bigg |_{\sigma=\sigma_0^+}\right)_{\rm M2-M5}&=&
-\frac{4\pi Q_2}{\Omega}\left[ 1-\frac{\sqrt{q_2}}{6}\left(1+\frac{27\sigma_0^4}{4\kappa^2 \Omega^2}\right)
\sigma_0^{-\frac{3}{2}}\beta^{-\frac{3}{2}}+\ldots) \right]~,
\nonumber\\
T_{M_2}&=&-\frac{4\pi Q_2}{\Omega}\left( 1-\frac{1}{3\cdot 2^{\frac{3}{5}}} q_2^{\frac{1}{5}}
\beta^{-\frac{6}{5}}+\ldots \right)
~.
\eea
The dots denote subleading contributions in the temperature. The right hand side of the first equation includes inside 
the parenthesis an $\Omega$-dependent contribution. Since we work in the limit of infinitesimal 
temperature and correspondingly infinitesimal $\sigma_0$ it is consistent to drop this term, $i.e.$ assume 
$\sigma_0^4 \ll \kappa^2 \Omega^2$.\footnote{In the opposite regime, 
$\sigma_0^4 \gg \kappa^2 \Omega^2$,
the matching (\ref{thermalda}) would give $\sigma_0\sim \beta^{\frac{3}{25}}$ which is inconsistent with 
the requirement $\sigma_0\ll \sigma_c$ in the small-temperature limit, $\beta \gg 1$, keeping $\Omega$
fixed.} Then, from the first matching condition in (\ref{thermalda}) we obtain
\beq\label{thermaldg}
\sigma_0^{(M)}=\left( \frac{q_2}{2^{4/3}\beta} \right)^{\frac{1}{5}}+\ldots
~.
\eeq
We are using the superscript $^{(M)}$ to keep track of the fact that this value is obtained from 
(\ref{thermalda}) by matching the tensions.

Similarly, from the expansion of angular densities we find
\bea\label{thermaldi}
\left( \frac{dJ}{dz}\bigg |_{\sigma=\sigma_0^+}\right)_{\rm M2-M5}&=&-\frac{2\pi Q_2}{\Omega^2}
\left[1-\sqrt{q_2}\left( \frac{2}{3}+\frac{9\sigma_0^4}{4\kappa^2 \Omega^2} \right)\sigma_0^{-\frac{3}{2}}
\beta^{-\frac{3}{2}}+\ldots \right]~,
\nonumber\\
\JJ_{M2}&=&-\frac{2\pi Q_2}{\Omega^2} \left( 1-\frac{5}{3\cdot 2^{\frac{8}{5}}} q_2^{\frac{1}{5}}
\beta^{-\frac{6}{5}}+\ldots\right)
~.
\eea
Dropping again the $\frac{\sigma_0^4}{\kappa^2 \Omega^2}$ term in the right hand side of the first 
equation we obtain
\beq\label{thermaldj}
\sigma_0^{(J)}=\frac{2^{\frac{13}{15}}}{5^{\frac{2}{3}}} \left( \frac{q_2}{\beta}\right)^{\frac{1}{5}}+\ldots
~.
\eeq
From the expansion of the entropy densities we obtain
\bea\label{thermaldk}
\left( \frac{dS}{dz}\bigg |_{\sigma=\sigma_0^+}\right)_{\rm M2-M5}=
-\frac{8\pi^2Q_2}{3\Omega}\sqrt{q_2} \sigma_0^{-\frac{3}{2}}\beta^{-\frac{1}{2}}+\ldots~,~~
s_{M2}=-\frac{4\pi^2 Q_2}{3\Omega} 2^{\frac{2}{5}} \left( \frac{q_2}{\beta} \right)^{\frac{1}{5}}+\ldots
\eea
yielding the matching point
\bea\label{thermaldl}
\sigma_0^{(S)}=\left( \frac{4q_2}{\beta} \right)^{\frac{1}{5}}+\ldots
~.
\eea
Finally, from the expansion of the rotation radius we find
\bea\label{thermaldm}
\rho(\sigma_0)_{\rm M2-M5}=\frac{1}{\Omega}\left( 1-\frac{\sqrt{q_2}}{2} \sigma_0^{-\frac{3}{2}} 
\beta^{-\frac{3}{2}}+\ldots \right)~,~~
\rho_{\rm M2}=\frac{1}{\Omega}\left( 1-\frac{1}{2^{\frac{8}{5}}} q_2^{\frac{1}{5}} \beta^{-\frac{6}{5}}+\ldots
\right)
\eea
and the same leading order expression for the matching point $\sigma_0^{(\rho)}$ as in (\ref{thermaldl}).

It is satisfying to verify the same scaling behavior (\ref{thermalcb}) from all the matching conditions and the 
corresponding hierarchy (\ref{thermalcc}). The numerical coefficients of the precise leading order behavior 
of $\sigma_0(\beta)$ are also comparable and of the same order 
\beq\label{thermaldo}
\frac{\sigma_0^{(J)}}{\sigma_0^{(M)}}
=\frac{2^{\frac{17}{15}}}{5^{\frac{2}{3}}}\simeq 0.750~, ~~
\frac{\sigma_0^{(\rho)}}{\sigma_0^{(S)}}=1~, ~~
\frac{\sigma_0^{(M)}}{\sigma_0^{(S)}}=2^{-2/3}\simeq0.630
~.
\eeq
We observe, however, that the degree of the matching varies according to the matched quantity and 
is less impressive than the corresponding 4\% matching in the static case 
\cite{Niarchos:2012cy}.\footnote{The apparent exact agreement between $\sigma_0^{(\rho)}$
and $\sigma_0^{(S)}$ is curious. We do not have a clear understanding of this fact.}
Presumably this is an effect of the absence of supersymmetry at zero temperature.

\vspace{0.2cm}
We can summarize our approximate construction of near-extremal rotating spikes as follows.
The leading order blackfold description is fixed by determining the radial dependence of the 
long-distance degrees of freedom $z(\sigma)$, $\rho(\sigma)$, $r_0(\sigma)$, $\alpha(\sigma)$
and $\theta(\sigma)$ with the use of eqs.\ (\ref{thermalag})-(\ref{thermalaj}), and (\ref{thermalbb}),
(\ref{thermalbc}) with 
\beq\label{thermaldp}
\sigma_0\simeq \left( \frac{\zeta q_2}{\beta} \right)^{\frac{1}{5}}+\ldots
~.
\eeq
$\zeta$ is an order one numerical constant that is approximately fixed by the 
above-described schemes.

\subsection{Near-extremal thermodynamics}
\label{btz}

The star of Ref.\ \cite{Niarchos:2012cy} was the near-extremal entropy of the static M2-M5 intersection
that gave rise to the intriguing central charge expression (\ref{introbd}), or equivalently (\ref{introbf}).
It is interesting to repeat that computation for the above stationary M2-M5-KKW solution. 

The entropy of the near-extremal spike has an infrared divergence as 
$\sigma \to +\infty$ that needs to be regularized (we refer the reader to Ref.\ \cite{Niarchos:2012cy} for
additional details). A renormalized version of the entropy gives
\bea\label{thermalea}
S&=&\frac{\Omega_{(3)}\Omega_{(4)} \pi \beta^4}{2G} \int_{\sigma_0}^{+\infty} d\sigma \, \sigma^3 
\rho \sqrt{1+{z'}^2+{\rho'}^2} \frac{(1-\Omega^2 \rho^2)^{\frac{3}{2}}}{\cosh^3 \alpha}
\nonumber\\
&\simeq& \frac{\pi \Omega_{(3)} \Omega_{(4)} q_5^{\frac{3}{2}}}{2G \Omega \sqrt \beta}
\int_{\sigma_0}^{+\infty} d\sigma \frac{(\kappa^2+\sigma^6)^{\frac{5}{4}}}
{\sigma^{\frac{3}{2}} \sqrt{\sigma^6 - \sigma_0^6}} +\OO(\beta^{-2})
\nonumber\\
&\simeq& \frac{\pi \Omega_{(3)} \Omega_{(4)} q_5^{\frac{3}{2}}}{2G \Omega \sqrt \beta}
\frac{\sigma_0^4 \Gamma\left(\frac{1}{3}\right) \Gamma\left(\frac{5}{6}\right)}{48\sqrt \pi}
~_2F_1\left( -\frac{5}{4},-\frac{2}{3},-\frac{1}{6}; -\frac{\kappa^2}{\sigma_0^6}\right) +\OO(\beta^{-2})
\nonumber\\
&\simeq& \frac{\pi^{\frac{3}{2}} \Omega_{(3)}\Omega_{(4)} q_5^{\frac{3}{2}}}
{G\Omega \sqrt \beta} \frac{\kappa^{\frac{5}{2}}}{\sigma_0^{\frac{7}{2}}} 
\frac{\Gamma\left( \frac{7}{12} \right)}{\Gamma\left(\frac{1}{12} \right)} +\Omega(\beta^{-2})
~.
\eea
In the last equality we picked the first term in the expansion $\sigma_0 \ll \kappa^{\frac{1}{3}}$.
Substituting the leading order expression (\ref{thermaldp}) for $\sigma_0$ we find (up to an overall 
numerical coefficient)
\beq\label{thermaleb}
S\sim \frac{1}{G} \frac{q_2^{\frac{9}{5}}}{\Omega} \frac{\beta^{\frac{1}{5}}}{q_5}
\eeq
which diverges in the zero temperature limit. Clearly, there is an important exchange of limits issue here
where taking the zero temperature limit first and then doing the integral gives a different result from 
doing the integral first and then taking the zero temperature limit.

An analogous computation of the angular momentum of the solution gives
\bea\label{thermalec}
J&=&\frac{3\Omega_{(3)}\Omega_{(4)}\Omega\beta^3}{8 G}\int_{\sigma_0}^{+\infty} d\sigma\sigma^{3}\rho^3\sqrt{1+z'^2+\rho'^2}\
\frac{\sqrt{1-\Omega^2\rho^2}}{\cosh^3\alpha} 
\nonumber\\
&\simeq& \frac{3\Omega_{(3)}\Omega_{(4)}q_5}{8G\Omega^2}\int_{\sigma_0}^{+\infty}
d\sigma\frac{\kappa^2+\sigma^6}{\sqrt{\sigma^6-\sigma_0^6}}+{\cal O}(\beta^{-\frac{3}{2}})
\nonumber\\
&\simeq& \frac{\Omega_{(3)}\Omega_{(4)}q_5}{G\Omega^2}\frac{\sigma_0^4\Gamma(1/3)\Gamma(5/6)}{128\sqrt{\pi}}
\left(1+\frac{4\kappa^2}{\sigma_0^6}\right)+{\cal O}(\beta^{-\frac{3}{2}})
\nonumber\\
&\simeq&\frac{\Omega_{(3)}\Omega_{(4)}q_5}{G\Omega^2}
\frac{\kappa^2}{\sigma_0^2} 
\frac{\Gamma(1/3)\Gamma(5/6)}{32\sqrt{\pi}}
+ \OO(\beta^{-\frac{3}{2}})
~.
\eea
In the last equality we picked again the first term in the expansion $\sigma_0 \ll \kappa^{\frac{1}{3}}$.

Combining the two expressions (\ref{thermalea}) and (\ref{thermalec}) we find a particularly 
simple expression for the entropy
\beq\label{thermaled}
S \simeq \frac{32 \pi^{\frac{1}{4}}}{\sqrt{6\zeta}} 
\frac{\Gamma\left( \frac{7}{12} \right)}{\Gamma\left( \frac{1}{12}\right) \sqrt{\Gamma\left(\frac{1}{6}\right)
\Gamma\left(\frac{1}{3} \right)}}
\sqrt{\frac{N_2^2 J}{N_5}}
+\ldots
~.
\eeq
Although this is the expression for the entropy of a {\it near-extremal} black hole, the leading term is 
reminiscent of the Bekenstein-Hawking entropy of an extremal rotating BTZ black hole
\beq\label{thermalee}
S_{\rm BTZ} =  2\pi \sqrt{\frac{c_{\rm BTZ} J}{6}}
\eeq
for a dual CFT with central charge $c_L=c_R=c_{\rm BTZ}$. A naive comparison with eq.\ (\ref{thermaled}) 
gives
\beq\label{thermalef}
c_{\rm BTZ} \simeq \{0.014, 0.436, 0.578\} \frac{N_2^2}{N_5}
\eeq
where each of the three numerical coefficients corresponds to $\zeta$ obtained respectively from the
matching scheme based on the entropy, mass and angular momentum densities. We observe that 
the $(N_2, N_5)$ scaling of this expression is exactly the same as the one deduced in \cite{Niarchos:2012cy} from the static
M2-M5 intersection (\ref{introbf}). It is amusing that even the numerical 
coefficient ($\sim 0.6$ in (\ref{introbf})) is comparable with the above coefficients ---especially the
third one based on the matching of angular moment densities in (\ref{thermaldj}). 

This agreement may be sensible since both in the static and stationary cases we are probing
the same two-dimensional superconformal field theory at the orthogonal M2-M5 intersection in 
different states. Presumably, in the static case we are probing a ground state with vanishing 
left-right scaling dimensions, whereas in the stationary rotating case we are probing a state
with the left-moving sector in an excited state. Despite these expectations, there is a lot that remains
to be understood about the validity of the overall picture, the precise physical interpretation of
(\ref{thermaled}) and the intriguing agreements between (\ref{thermalef}) and (\ref{introbf}).
We comment further on this aspect in the next concluding section.

\section{Open problems}
\label{conclude}

In Refs.\ \cite{Niarchos:2012pn,Niarchos:2012cy} and the above presentation we discussed the leading 
order effective field theory description of fully localized orthogonal M2-M5 intersections. It should be 
appreciated that the exact supergravity solutions of these systems (both extremal and non-extremal) are not 
known and therefore the effective descriptions of this work are opening the route to a new more 
efficient treatment. One of the primary ultimate goals in this application to the M2-M5 system is 
the identification of the two-dimensional $\NN=(4,4)$ SCFT at the intersection by means of holography
and its implications for the microscopic structure of the M5 brane in M-theory. Preliminary results like
the formulae (\ref{introbd}) are indicative of a rich underlying structure that remains to be explored.

In much of this report we described the assumptions, limitations, but also unexpected merits of a leading 
order effective field theory treatment of the intersection. An important part of the story, and the natural
next step in the analysis we have performed, is the development of the precise map between a solution
of the hydrodynamic effective blackfold theory and the corresponding bulk supergravity solution. 
Much like in the fluid-gravity correspondence it is expected that this map is one-to-one. Considerable 
progress has been achieved in proving this fact in the fluid-gravity version of the AdS/CFT correspondence, 
however it remains a largely open and rather involved technical problem in the case of general 
blackfolds. So far the state-of-the-art in this direction has been achieved in non-supersymmetric pure gravity
contexts \cite{Emparan:2007wm,Emparan:2009vd,Camps:2012hw}. Further technical progress in general 
supergravity theories appears to be most imminent for extremal configurations \cite{NiarchosSiampos}.

For the M2-M5 system, which is the system of interest in this work, a direct reconstruction of the bulk
supergravity solution will be useful for several reasons, some of which can be summarized as follows:
\begin{itemize}
\item For static $1/4$-BPS configurations it will allow to make contact with the exact solution generating
techniques of Ref.\ \cite{Lunin:2007mj}. Identifying the near-horizon geometry will be a big step forward in 
developing the $AdS_3/CFT_2$ correspondence of this system and verifying the validity of the central
charge expressions (\ref{introbd}). In particular, it will allow us to verify the applicability of the 
general central charge formula (\ref{introab}) and determine the precise relation between
the integers $k_+$, $k_-$ and the number of M2 and M5 branes $N_2, N_5$ respectively.
It is also interesting to explore whether the new expansion in powers of
$\frac{1}{\lambda}$ (see eq.\ (\ref{introba})) simplifies this analysis and takes us one step further compared
to previous work. In addition, knowledge of the exact supergravity solution will help clarify the convergence
properties of the perturbative blackfold expansion scheme and the miraculous agreements observed
in Ref.\ \cite{Niarchos:2012pn}.
\item The corresponding analysis of the extremal non-supersymmetric orthogonal M2-M5-KKW 
intersection will help clarify analogous questions in a non-supersymmetric setting, in particular the 
mysteries of section \ref{btz}.
\item More generally, the development of a precise one-to-one map between effective hydrodynamic
solutions and supergravity backgrounds will be very useful in clarifying the potential holographic 
nature of such maps where one links a supergravity solution to an effective hydrodynamic (blackfold)
solution to a microscopic state of the underlying brane system. This is potentially a promising new 
route towards a much more general understanding of holography beyond the near-horizon limit and
anti-deSitter spaces. The recent work \cite{Caldarelli:2012hy} is based on a related logic.
\end{itemize}

We hope to return with a detailed discussion of these issues in future work.

\section*{Acknowledgements}

We would like to thank Jan de Boer, Jerome Gauntlett, Elias Kiritsis, Niels Obers, Kostas Skenderis and 
Arkady Tseytlin for useful comments on the literature and inspiring discussions.
The research of VN was in part supported by grants 
PERG07-GA-2010-268246, and the EU program ``Thales'' ESF/NSRF 2007-2013. It was also 
co-financed by the European Union (European Social Fund, ESF) and Greek national
funds through the Operational Program ``Education and Lifelong Learning'' of the National
Strategic Reference Framework (NSRF) under ``Funding of proposals that have received a
positive evaluation in the 3rd and 4th Call of ERC Grant Schemes''.
The research of KS has been supported by an ARC contract No.\ AUWB-2010-10/15-UMONS-1, 
a Fund for Scientific Research-FNRS (Belgium), by IISN-Belgium (convention 
4.4511.06), by 
"Communaut\'e fran\c{c}aise de Belgique - Actions de Recherche Concert\'ees",
the ITN programme PITN-GA-2009-237920, the ERC Advanced Grant 226371, the IFCPAR CEFIPRA 
programme 4104-2 and the ANR programme blanc NT09-573739. KS would like also
to thank the University of Patras and the University of Crete for hospitality where part of this work was done.

\appendix

\section{Thermodynamics of stationary M2 brane cylinders}
\label{mthermo}

For quick reference in this appendix we list the properties of a rotating M2 black cylinder. 

We consider an M2 black brane in flat space parametrized in cylindrical coordinates as
\beq\label{maa}
ds^2=-dt^2+\rho^2 d\theta^2+d\rho^2+dz^2+\sum_{i=1}^7 (dx^i)^2
~.
\eeq
The M2 brane is oriented along the directions $(t,\theta,z)$ and rotates along the angular 
$\theta$ direction with velocity vector 
\beq\label{mab}
u=\frac{1}{\sqrt{1-\Omega^2 \rho^2}}\left( \frac{\d}{\d t}+\Omega \frac{\d}{\d\theta} \right)
~.
\eeq
The relevant thermodynamic quantities 
\beq\label{mac}
\varepsilon=\frac{\Omega_{(7)}}{16\pi G}r_0^6 (1+6 \cosh^2 \alpha)~, ~~
\TT=\frac{3}{2\pi r_0 \cosh\alpha}~, ~~ 
s=\frac{\Omega_{(7)}}{4G}r_0^7 \cosh\alpha
~,
\eeq
\beq\label{mad}
Q_2=-\frac{\Omega_{(7)}}{16\pi G} 6r_0^6 \sinh\alpha \cosh\alpha
~, ~~ \Phi_2=-\tanh\alpha
\eeq
imply the stress-energy tensor
\beq\label{mae}
T_{ab}=\TT s \left (u_a u_b-\frac{1}{6} \gamma_{ab} \right) -\Phi_2 Q_2 \gamma_{ab}
~.
\eeq

At equilibrium the unknown worldvolume functions $r_0,\alpha, \rho$ can be determined in 
terms of the constants $T,Q_2,\Omega$ using the equations
\beq\label{maf}
\tilde q_2=-\frac{16\pi G}{6\Omega_{(7)}}Q_2=r_0^6 \sinh\alpha \cosh\alpha
~,
\eeq
\beq\label{mag}
r_0\cosh\alpha=\sqrt{1-\Omega^2 \rho^2}\, \tilde \beta~, ~~
\tilde \beta:=\frac{3}{2\pi T}
~,
\eeq
\beq\label{mai}
\Omega^2 \rho^2=\frac{1+6\sinh^2\alpha}{1+6\cosh^2 \alpha}
~.
\eeq
Comparing (\ref{maf}), (\ref{mag}) with (\ref{thermalad}),(\ref{thermalaf}) we can easily find that $(\tilde q_2,\tilde\beta)=(8q_2\, ,2\beta)$.  
Notice that (\ref{mai}) is consistent with equation (3.6) of \cite{Emparan:2011hg} 
for $p=1$ and it can be also derived by requiring $T^{11}=0$.
It is useful to rewrite eqs.\ (\ref{maf}), (\ref{mag}) as\\

\beq\label{mao}
r_0=\left(\frac{\tilde q_2}{\sinh\alpha\cosh\alpha}\right)^{1/6}~,~~
\tilde\beta
=\cosh\alpha\left(\frac{\tilde q_2}{\sinh\alpha\cosh\alpha}\right)^{1/6}\sqrt{\frac{1+6\cosh^2\alpha}{6}}
\eeq
where use was made of (\ref{mai}).  

Accordingly, the thermodynamics of the general non-extremal cylinder solution is
\beq\label{maj}
M=-\frac{4\pi Q_2}{3\Omega}L_z\sqrt{\frac{1+6\sinh^2\alpha}{1+6\cosh^2\alpha}}\ \frac{2+3\sinh^2\alpha}{\sinh\alpha\cosh\alpha}
~,
\eeq
\beq\label{mak}
J=-\frac{\pi Q_2}{3\Omega^2}L_z\frac{(1+6\sinh^2\alpha)^{3/2}}{\sqrt{1+6\cosh^2\alpha}}\ \frac{1}{\sinh\alpha\cosh\alpha}
~,
\eeq
\beq\label{mal}
S=-\frac{(2\pi)^2Q_2}{3\Omega}L_z r_0\sqrt{\frac{1+6\sinh^2\alpha}{1+6\cosh^2\alpha}}\,\frac{\tilde\beta}{\sinh\alpha\cosh\alpha}
~
\eeq

In the extremal limit, where $(Q_2,\Omega)$ are kept fixed, eqs.\ (\ref{mai}), (\ref{mao}) imply
\beq\label{map}
\tilde \beta\to\infty\ ,\qquad \alpha\to\infty\ ,\qquad r_0\to0\ ,\qquad \rho=\frac{1}{\Omega}
~.
\eeq
Moreover, using the definition (\ref{extremeaa}) one can check that the null momentum density is
\beq\label{maq}
\KK=|Q_2|
\eeq
and the rest of the thermodynamics becomes
\beq\label{mar}
\frac{M}{L_z}=-\frac{4\pi Q_2}{\Omega}~,~~
\frac{J}{L_z}=-\frac{2\pi Q_2}{\Omega^2}~,~~
\frac{S}{L_z}=-\frac{(2\pi)^2Q_2}{3\Omega}\ r_0\to 0
~.
\eeq

Away from extremality we have to solve (\ref{mao}) for non-zero temperatures.
This can be done easily with a perturbative expansion
\beq\label{mat}
\cosh\alpha=y^{3/5}-\frac{1}{10\, y^{3/5}}-\frac{43}{1200\, y^{9/5}}
-\frac{137}{4320\, y^{3}}+{\cal O}(y^{-21/5})~,~~
y:=\frac{\tilde\beta}{\tilde q_2^{1/6}}
\eeq
around extremality.
Plugging this expansion into the expressions (\ref{mai}), (\ref{maj}), (\ref{mak}), (\ref{mal}) 
we find respectively 
\beq\label{mau}
\begin{array}{c c}
&\rho\simeq\frac{1}{\Omega}\left(1-\frac{\tilde q_2^{1/5}}
{2\tilde \beta^{6/5}}-\frac{17\tilde q_2^{2/5}}{120\tilde \beta^{12/5}}+{\cal O}(y^{-18/5})\right)~,
\nonumber\\
&M\simeq-\frac{4\pi Q_2}{\Omega}L_z\left(1-\frac{\tilde q_2^{1/5}}
{3\tilde \beta^{6/5}}+\frac{\tilde q_2^{2/5}}{60\tilde \beta^{12/5}}+{\cal O}(y^{-18/5})\right)~,
\nonumber\\
&J\simeq-\frac{2\pi Q_2}{\Omega^2}L_z\left(1-\frac{5\tilde q_2^{1/5}}{6\tilde \beta^{6/5}}-
\frac{\tilde q_2^{2/5}}{12\tilde \beta^{12/5}}+{\cal O}(y^{-18/5})\right)~,
\nonumber\\
&S\simeq-\frac{(2\pi)^2 Q_2}{3\Omega}L_z\left(\frac{\tilde q_2^{1/5}}{\tilde \beta^{1/5}}
+\frac{\tilde q_2^{2/5}}{5\tilde \beta^{7/5}}+\frac{37\tilde q_2^{3/5}}{200\tilde \beta^{13/5}}
+{\cal O}(y^{-23/5})\right)
~.
\end{array}
\eeq

\section{Extrinsic curvature and stress-energy tensors}
\label{extcurvature}

In this appendix we provide the expressions of the extrinsic curvature and stress-energy tensors
that are relevant for sections \ref{extremespikes} and \ref{thermalspikes}.

The extrinsic curvature of the embedding (\ref{ansbab}) can be found easily with the use of equation (A.20) 
in \cite{Emparan:2009at}. The non-vanishing components take the form
\beq\label{ringav}
\begin{array}{c c}
&K_{11}{}^z=\frac{\rho z'\rho'}{1+z'^2+\rho'^2}~,~~ 
K_{22}{}^{z}=\frac{(1+\rho'^2)z''-\rho'z'\rho''}{1+z'^2+\rho'^2}~,~~ 
K_{33}{}^{z}=\frac{\sigma z'}{1+z'^2+\rho'^2}~,~~
K_{44}{}^{z}=\frac{\sigma z'\sin^2\psi}{1+z'^2+\rho'^2}~,  
\nonumber\\
&K_{11}{}^{\rho}=-\frac{\rho(1+z'^2)}{1+z'^2+\rho'^2},~
K_{22}{}^{\rho}=\frac{(1+z'^2)\rho''-\rho'z'z''}{1+z'^2+\rho'^2},~ 
K_{33}{}^{\rho}=\frac{\sigma \rho'}{1+z'^2+\rho'^2},~
K_{44}{}^{\rho}=\frac{\sigma \rho'\sin^2\psi}{1+z'^2+\rho'^2}, 
\nonumber\\
&K_{11}{}^r=\frac{\rho\rho'}{1+z'^2+\rho'^2},~ 
K_{22}{}^r=-\frac{z'z''+\rho'\rho''}{1+z'^2+\rho'^2},~
K_{33}{}^r=-\sigma\frac{z'^2+\rho'^2}{1+z'^2+\rho'^2},~ 
K_{44}{}^r=-\sigma\sin^2\psi\frac{z'^2+\rho'^2}{1+z'^2+\rho'^2},
\nonumber\\
&K_{55}{}^{z}=\frac{\sigma z'\sin^2\psi\sin^2\phi}{1+z'^2+\rho'^2}~,~~ 
K_{55}{}^{\rho}=\frac{\sigma \rho'\sin^2\psi\sin^2\phi}{1+z'^2+\rho'^2}~,~~
K_{55}{}^{r}=-\sigma\sin^2\psi\sin^2\phi\frac{z'^2+\rho'^2}{1+z'^2+\rho'^2}~,
\nonumber\\
\end{array}
\eeq
Accordingly the mean curvature vector $K^\mu=\gamma^{ab}K_{ab}{}^\mu$ has three non-vanishing 
components
\bea\label{ringaw}
&&K^r=\frac{\rho'}{\rho(1+z'^2+\rho'^2)}
-\frac{z'z''+\rho'\rho''}{(1+z'^2+\rho'^2)^2}-\frac{3(z'^2+\rho'^2)}{\sigma(1+z'^2+\rho'^2)}~,
\nonumber\\
&&K^z=\frac{z'\rho'}{\rho(1+z'^2+\rho'^2)}
+\frac{(1+\rho'^2)z''-\rho'z'\rho''}{(1+z'^2+\rho'^2)^2}+\frac{3z'}{\sigma(1+z'^2+\rho'^2)}~,
\\
&&K^\rho=-\frac{1+z'^2}{\rho(1+z'^2+\rho'^2)}
+\frac{(1+z'^2)\rho''-\rho'z'z''}{(1+z'^2+\rho'^2)^2}+\frac{3\rho'}{\sigma(1+z'^2+\rho'^2)}
~.\nonumber
\eea

Characteristic curvature sizes of a configuration can be defined as $L^{(i)}_{\rm curv}=|K_{(i)}|^{-1}$
with $K_{(i)}=K^\mu n^{(i)}{}_\mu$. $n^{(i)}{}_\mu\ ,i=z,\rho$, are the unit normal vectors of the embedding 
surface of the fivebrane, $z=z(\sigma)$, $\rho=\rho(\sigma)$ respectively,
\beq\label{ringaw}
n^{(z)}{}_\mu=\frac{1}{\sqrt{1+z'^2}}\ (\vec{0}_2,-z',\vec{0}_3,1,\vec{0}_4)\ ,\qquad 
n^{(\rho)}{}_\mu=\frac{1}{\sqrt{1+\rho'^2}}\ (\vec{0}_2,-\rho',\vec{0}_4,1,\vec{0}_3) 
~.
\eeq
After some algebra we compute the principal curvatures
\beq\label{ringawa}
K_{(z)}=\frac{1}{\sqrt{1+z'^2}}\left(\frac{z''}{1+z'^2+\rho'^2}+\frac{3z'}{\sigma}\right)~,~~
K_{(\rho)}=\frac{1}{\sqrt{1+\rho'^2}}\left(\frac{\rho''}{1+z'^2+\rho'^2}+\frac{3\rho'}{\sigma}-\frac{1}{\rho}\right)
~.
\eeq
The mean curvature equals the half of the sum of the principal curvatures. 

Finally, the stress-energy tensor (\ref{eqsad}) that enters into the extrinsic equations of 
motion (\ref{eqsba}) and the thermodynamics of the solution (\ref{eqscb}), has 
the non-vanishing components
\bea\label{ringax}
&&T^{00}=\frac{{\cal I}}{1-\Omega^2\rho^2}(1+3\cosh^2\alpha-\Omega^2\rho^2(1+3\sinh^2\alpha))~,~~ 
\nonumber\\
&&T^{11}=-\frac{{\cal I}(1+3\sinh^2\alpha)}{\rho^2}+{\cal I}\frac{3\Omega^2}{1-\Omega^2\rho^2}~,~~
T^{22}=-\gamma^{22}{\cal I}(1+3\sinh^2\alpha)~,
\\ 
&&T^{33}=-\gamma^{33}{\cal I} {\cal J}~, ~~
T^{44}=-\gamma^{44}{\cal I}  {\cal J}~,~~
T^{55}=-\gamma^{55}{\cal I}  {\cal J}~,~~
T^{01}=T^{10}={\cal I}\frac{3\Omega}{1-\Omega^2\rho^2}~,
\nonumber
\eea
where 
\beq
{\cal I}:=\frac{r_0^3\Omega_{(4)}}{16\pi G}~,~~
{\cal J}:=1+3\cos^2\theta\ \sinh^2\alpha
~.
\eeq

\end{document}